\documentclass[sigconf]{acmart}

\usepackage{graphicx}
\usepackage{subfigure}
\usepackage{subcaption}

\usepackage{tikz}
\usetikzlibrary{shapes.geometric, arrows}
\usetikzlibrary{calc,arrows,shapes,positioning}
\usepackage{tikz-cd}
\usetikzlibrary{arrows,decorations.pathmorphing,backgrounds,positioning,fit,matrix}
\usepackage{booktabs,caption,fixltx2e}
\usepackage[flushleft]{threeparttable}
\usepackage[percent]{overpic}
\usepackage{xcolor}
\usepackage{subcaption}
\tikzset{>=latex}

\setcopyright{none}
\renewcommand\footnotetextcopyrightpermission[1]{} % removes footnote with conference information in first column

\begin{document}

%\title[Yahoo Ad Exchange: Optimizing Floors in First Price Auction]{Yahoo Ad Exchange: Optimizing Floors in First Price Auctions \\\vspace{1em}
%  \Large Miguel Alcobendas, Amado Diaz, Oriol Diaz, Hermakumar Gokulakannan, Jonathan Ji, Boris Kapchits, Rabi Kavoori, Maria Rosario Levy Roman, Emilien Pouradier-Duteil, Korby Satow, Swarna Veerapaneni, Dawit Wami \\%\vspace{1em}
%  \normalfont \{lisbona,amado.diaz,oriold,hemakumar.gokulakannan,jonathan.ji,boris.kapchits,rabi.gubbala,\\maria.rosario.levy.roman,emilien.pouradier,korbys,swarna.veerapaneni,dawit.wami\}@yahooinc.com \\\vspace{1em}
%  \Huge Yahoo
%  }
%\title{Optimizing Floors in First Price Auctions: an Empirical Study of Yahoo}
%\author{Miguel Alcobendas\\Yahoo Research \and Shunto J. Kobayashi\\Caltech \and Ke Shi\\Caltech \and  Matthew Shum\\Caltech}

\title{Optimizing Floors in First Price Auctions: an Empirical Study of Yahoo Advertising
\\\vspace{1em}
 \Large Miguel Alcobendas, Amado Diaz, Oriol Diaz, Hemakumar Gokulakannan, Jonathan Ji, Boris Kapchits, Rabi Kavoori, Maria Rosario Levy Roman, Korby Satow, Swarna Veerapaneni, Dawit Wami \\%\vspace{1em}
\normalfont \{lisbona,amado.diaz,oriold,hemakumar.gokulakannan,jonathan.ji,boris.kapchits,rabi.gubbala,\\maria.rosario.levy.roman,korbys,swarna.veerapaneni,dawit.wami\}@yahooinc.com \\\vspace{0.5em}
\Huge Yahoo \\\vspace{0.5em}
\Large \bfseries Emilien Pouradier Duteil\\%\vspace{1em}
\normalfont emilien.pouradierduteil@gmail.com \\\vspace{0.5em}
\Huge Moloco
}

%\maketitle

\begin{abstract}
%\vspace*{-1.5em}
Floors (also known as reserve prices) help publishers to increase the expected revenue of their ad space, which is usually sold via auctions. Floors are defined as the minimum bid that the a seller (it can be a publisher or an ad exchange) is willing to accept for the inventory opportunity. In this paper, we present a model to set floors in first price auctions, and discuss the impact of its implementation on Yahoo sites. The model captures important characteristics of the online advertising industry. For instance, some bidders impose restrictions on how ad exchanges can handle data from bidders, conditioning the model choice to set reserve prices. Our solution induces bidders to change their bidding behavior as a response to the floors enclosed in the bid request, helping online publishers to increase their ad revenue. 

The outlined methodology has been implemented at Yahoo with remarkable results. The annualized incremental revenue is estimated at +1.3\% on Yahoo display inventory, and +2.5\% on video ad inventory. These are non-negligible numbers in the multi-million Yahoo ad business.

\end{abstract}

\settopmatter{printfolios=true}
\maketitle
\pagestyle{plain}

\section{Introduction}\normalsize

 Floors (also known as reserve prices) help publishers to increase the expected revenue of their ad space, which is usually sold via auctions. Floors are defined as the minimum bid that the a seller (it can be a publisher or an ad exchange) is willing to accept for the inventory opportunity. The approach for setting floors may vary depending on the auction design. In this paper, we present a model to set floors and discuss the impact of its implementation on Yahoo sites. We focus on a first price auction mechanism, the current most popular approach to price and allocate online display advertising. This was not the case a few years ago, when the second price rule was the most popular mechanism. For instance, the Yahoo's display exchange transitioned from a second to a first price auction rule in 2019 (see \cite{alcobendas22}).  In first price auctions, the highest bid wins (conditional on being greater than the reserve price) and the winner pays its bid. This is different from the second price mechanism, where the highest bid also wins the auction but the winner pays the maximum of the second highest bid and the reserve price.

The model captures important characteristics of the online advertising industry. For instance, some bidders impose restrictions on how ad exchanges can handle data from bidders. As we will discuss in detail later on, these restrictions condition the model choice to set reserve prices. The algorithm discussed in this paper accommodates two typical constraints imposed by large bidders: first, exchanges cannot use intraday bid data to compute floors. Second, for a given ad opportunity, the floor value has to be the same among bidders within the same type. In this paper we consider two types of bidders: \textit{regulars} and \textit{rebroadcasters}. \textit{Regular} bidders correspond to potential buyers directly bidding in the ad exchange on behalf of their advertising clients. On the other hand, \textit{rebroadcasters} are other exchanges bidding on the exchange. As the name implies, they "rebroadcast" the advertising opportunities to their own bidders. The second constraint implies that \textit{regular} bidders face the same floor (same for \textit{Rebroadcasters}).

The outlined methodology has been implemented at Yahoo with remarkable results.
In June 2021, we deployed the model in production on display ad inventory in Yahoo websites located in North-America. Afterwards, we rolled out the feature in other markets, and in October 2022 we started optimizing floors on Yahoo video ad inventory. This product significantly improves the previous approach to set floors: a manual process that did not take into account the impact of floors on the bidding behavior of buyers. The annualized incremental revenue is estimated at +1.3\% on Yahoo display inventory and +2.5\% on video ad inventory. These are non-negligible numbers in the multi-million Yahoo ad business.

The rest of the paper is organized as follows. Section 2 summarizes previous literature. Section 3 highlights details about the impact of reserve prices in first price auctions. Section 4 describes the model. Section 5 includes details about the implementation of the algorithm at Yahoo. Finally, section 6 concludes.

\section{Previous Literature}

There are not many empirical studies documenting controlled experiments on reserve prices. \cite{ostrovsky23} is one of the first large scale experiments measuring the impact of setting floors on Yahoo sponsored search auctions, which used a generalized second price auction mechanism to allocate and price advertisements. \cite{alcobendas16} uses the classical approaches developed by \cite{riley81} and \cite{myerson81} to set floors in online display advertising when Yahoo used a second price rule. However, as discussed in detail later on, the mechanics governing the impact of floors on first price auctions are very different. \cite{reiley06} conducted an experiment to investigate the impact of reserve prices in a first price online auction for trading cards. In line with auction theory, he found that floors reduce the probability of a sale and increase bids when they are present. We observed similar results in our study. Our work differs in that we optimize floors at scale and focus on online advertising, a multi-billion dollar business.

There are several studies proposing different algorithms to set reserve prices. Some of them include an offline analysis with real auction data to validate results. However, they do not report details of a large scale implementation. There exists a vast algorithmic literature about second price auctions (e.g. \cite{Paes16}, \cite{mohri16}, and \cite{drutsa20}). However, there is no guarantee that floors optimized for second price auctions are effective in a first price environment. There are less work on first price. For instance, \cite{feng21} proposed an algorithm to set floors in first price auctions using a gradient-based algorithm. Our approach differs in several aspects: 1) We enclose different reserve price values in the bid request depending on the bidder's type, 2) Contractual agreements with large bidders typically prevent ad exchanges from using intraday bid data to set floors. As a result, online feedback algorithms cannot be used to adjust floors during the day.

\section{Floors in First Price Auctions}

In this section we review the basic elements of an auction in online advertising, and discuss how floors may change the bidding behavior of the buyers, known as Demand Side Platforms (DSPs), increasing platform's revenue. DSPs are demand aggregators that bid on behalf of their advertiser clients.

\begin{figure}[htbp]
	\caption{Online Advertising Auction}
	\label{fig::ssp_auction}
	\centering
	\includegraphics[width=.35\textwidth]{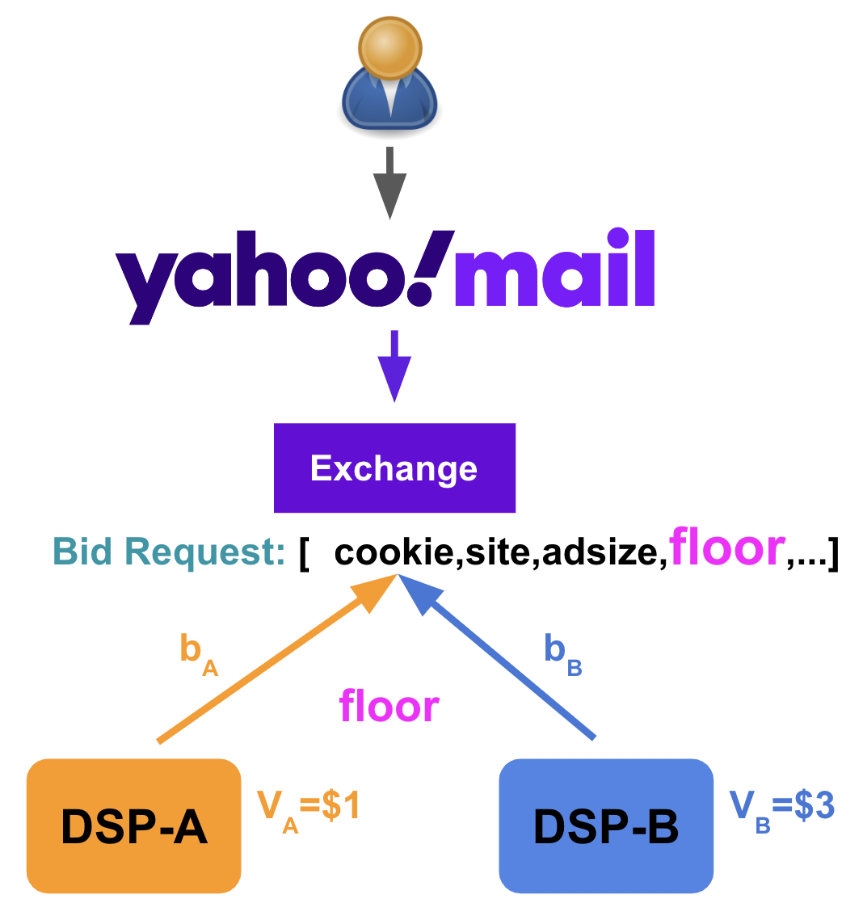}
	
	\medskip
	\small
    \bfseries Note: \normalfont stylized example to understand how the bidding behavior of DSPs changes as a function of the auction mechanism and floors.
    \medskip
\end{figure}

We use Figure~\ref{fig::ssp_auction} to illustrate how auctions work depending on how the winner and clearing price are determined: Second vs First price rules. As depicted in the figure, imagine that a user visits the Yahoo mail property. His visit triggers an auction to display an advertisement. The Exchange manages the auction and sends bid requests to the Demand Side Platforms (DSPs). In our example we have two potential bidders: DSP-A and DSP-B. The request has information that helps potential buyers to value the inventory opportunity and find the most appropriate advertisement. For instance, it may contain a cookie or device identifiers that help DSPs to match the visitor with their user profile databases. It may also include details about the browser (e.g. Chrome, Safari, Edge), URL, permitted advertisement size, time of the day, etc. Moreover, the bid request also contains the reserve price for that inventory opportunity (denoted as $floor$ in Figure~\ref{fig::ssp_auction}). As mentioned above, the floor is the minimum bid that a seller is willing to accept for the opportunity. Bidders evaluate the ad request, match the attached information with their own data sets, determine the value of the inventory opportunity, find an appropriate advertisement, and finally submit a bid. Assume that DSP-A values the request at \$1 ($V_A=\$1$) and DSP-B at \$3 ($V_B=\$3$). This valuation corresponds to the maximum amount that a potential buyer is willing to pay for the opportunity. Given the valuation, DSPs decide how much they are going to bid: DSP-A's bid is denoted as $b_A$ and DSP-B's bid as $b_B$. 

The auction design determines the bid response of DSPs. In order to understand the impact of floors on the bidding behavior of buyers participating in a first price mechanism, it is useful to first discuss the role of floors in a second price environment. As an historical note, most of the largest ad exchanges switched from second to a first price auction mechanism between 2018 and 2020. 

\subsection{Second Price Auction}
Assuming a simple second price auction without budget constraints, the winner is the bidder submitting the highest bid (conditional on being greater than the floor), and the clearing price is the maximum of the second highest bid and the floor. In such a scenario, each DSP has incentives to submit its own valuation independently on other factors like floors or expected competition from other buyers. Coming back to our example in Figure~\ref{fig::ssp_auction}, assume that the exchange sets a floor of 50 cents ($floor=0.5$). Then in a second price auction, DSP-B submits a bid of \$3 ($b_B=\$3$) and DSP-A's offer equals one dollar ($b_A=\$1$). DSP-B wins and pays the maximum of the floor and the second highest bid. Since the floor, set at \$0.5, is lower than the bid submitted by DSP-A, the winner pays the exchange \$1. The DSP-B's surplus, defined as the difference between its inventory valuation and the payment, equals \$2. If, for instance, we increase the $floor$ enclosed in the ad request to \$2.5, DSP-B still submits the same bid (\$3) and wins the auction, but it pays the floor value instead of the bid submitted by DSP-A (i.e. \$2.5 instead of \$1). DSP-B's surplus is still positive (\$3 - \$2.5 = \$0.5 ) and the exchange improves the monetization of the opportunity by receiving \$2.5 instead of \$1.

\subsection{First Price Auction}
In first price auctions the winner is the one submitting the highest bid (conditional on being above the floor) and pays its bid. The clearing price rule creates incentives for potential buyers to take into account both expected competition and the floor value when deciding their bids, since bidding their valuation is no longer the optimal strategy (see \cite{krishna10}).

In a first price auction environment and given the scenario depicted in Figure~\ref{fig::ssp_auction}, assume that DSP-A and DSP-B bid their valuations (i.e. $b_A=\$1$ and $b_B=\$3$), and the floor equals 50 cents ($floor=\$0.5$). DSP-B still wins the auction. However, it pays \$3 instead of the \$1 it would pay if the marketplace ran in a second price environment. As a result, DSP-B's surplus decreases. Now DSP-B's surplus equals \$0 instead of \$2. What would be the consequences of DSP-B bidding \$2 instead of \$3? It would still win the auction since DSP-A would never bid more than its valuation (\$1), and DSP-B's surplus would increase from \$0 to \$1.  So, depending on the level of competition, DSPs may be better off by submitting a bid lower than their valuation. In the literature this is called bid shading. For further details, \cite{gligorijevic20} propose a machine learning approach to compute the optimal bid shading using data from Yahoo DSP.

What would be the impact of floors on bidding behavior in a first price auction? Imagine now that DSP-A and DSP-B bid $b_A=\$1$ and $b_B=\$2$ respectively, and the floor sent to potential buyers equals \$2.5 ($floor=\$2.5$). Since all submitted bids are below the floor, the inventory opportunity is not allocated to any bidder. Hence bidders' surplus equals zero and the publisher does not get paid for the inventory opportunity. DSP-B would be better off if it increased its bid to \$2.5, since it would win the auction and its surplus would increase to \$0.5. At the same time, the publisher would monetize the inventory opportunity and receive \$2.5. The example illustrates how floors may change the bidding behavior of potential buyers, and how it can be used as a tool to increase publishers' revenue in a first price auction. 

Potential buyers interact with ad exchanges millions of times every day. DSPs decide how much they value each opportunity and the corresponding bid based on the information they have about the user, the expected probability of winning the auction, and the observed floor. Ex-ante, DSPs do not exactly know how much competition they are going to face for a particular auction. However, using historical data, DSPs have estimates about the probability of winning the impression as a function of the submitted bid, reserve prices, and other auction and user characteristics.

\begin{figure}[htbp]
	\caption{Bidding Distribution of a DSP}
	\label{fig::biddingdistribution}
	\centering
	\includegraphics[width=.450\textwidth]{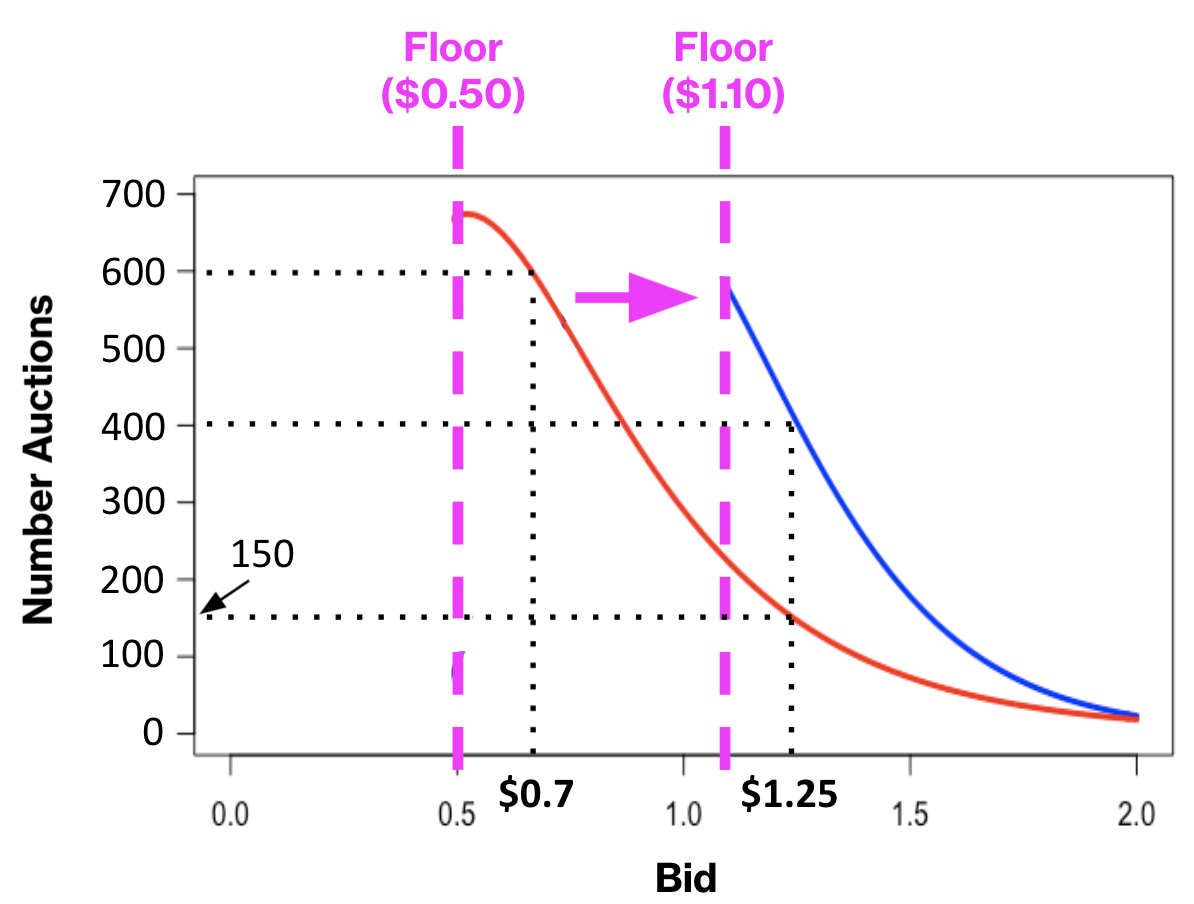}
	\medskip
    \small
    
    \bfseries Note: \normalfont Figure illustrates how the bidding distribution of a particular DSP shifts when increasing floors.
    \medskip
\end{figure}

Generalizing the results outlined above to multiple auctions, Figure~\ref{fig::biddingdistribution} illustrates how the bidding distribution of a particular DSP may change as a result of the floor included in the ad request when engaging with a particular exchange repeatedly. The x-axis displays the bid in dollars per 1,000 requests, and the vertical axis shows the number of submitted auctions given a particular bid. The red curve corresponds to the bidding distribution if the enclosed floor in the bid request equals 50 cents. For instance, the plot shows that the DSP participates in 600 auctions with a bid of 70 cents. Since the probability of winning the auction for bids below the floor is equal to zero, we assume that the DSP does not respond with bids below 50 cents. That explains why the distribution is truncated below the floor\footnote{Even-though the probability of winning an ad impression is zero for bids below the floor, in our implementation at Yahoo we observe that some bidders respond with bids below the floor value.}. Following the arguments stated above in the single auction case, DSP bid distributions may change as a result of changes in floors. The blue curve in Figure~\ref{fig::biddingdistribution} corresponds to the bid distribution of the same DSP bidding on the same inventory if the floor set by the exchange is set to \$1.10 instead of 50 cents. Following the above discussion, the DSP may find it profitable to raise its bids as long as they are lower than its corresponding inventory valuation. That is why increasing floors shifts the bid distribution to the right. At the same time, higher floors reduce the auction participation of DSPs because the valuation of some ad requests will lie between the original floor and the new higher one, and, as mentioned before, a potential buyer has no incentives to bid above its valuation. To illustrate this point, Figure~\ref{fig::biddingdistribution} shows that when the floor is set at \$1.10 (blue curve), the DSP participates in 400 auctions when the submitted bid equals \$1.25, while we only observe 150 auctions with the same bid value when the floor equals \$0.5 (red curve).

The seller wants to set floors that maximize its returns. If the floor set for a particular buyer is too low, the participation rate of the DSP may be very high, but the average clearing price of the auction may be too small. On the other hand, if the floor is set too high, the DSP participation may be too low to compensate the increase in the number of submitted high bids. Our algorithm exploits this trade-off to find the floors that maximize the expected return of the auction.

The mechanics governing the first price auction in online display advertising industry are similar to the one highlighted in the above illustration (Figure~\ref{fig::ssp_auction}). However, in the industry there is an extra component that must be considered, and it is the existence of an outside option (AKA waterfall). The outside option allows publishers to have another chance to monetize the ad opportunity when the main exchange fails to find a buyer. It can be implemented in several ways. For instance, if all submitted bids in the main exchange are below the floor, the inventory opportunity can be sent to another exchange to try to monetize the opportunity. This is the option adopted by Yahoo. Another popular approach is the main exchange to send a new bid request with a lower floor.

\vspace{2em}
\section{Model}
In this section we discuss the methodology to compute reserve prices that maximize the expected revenue of an online Publisher (e.g. Yahoo).

Our model assumes that a publisher serves ads using a first price mechanism with an outside option. As in the case of Yahoo, if all submitted bids in the main ad exchange are below the floor, the inventory opportunity is sent to another exchange (outside option). We rule out cases where the main ad exchange may have a different objective other than maximizing Publisher's revenue. In our application there is no such a problem because the exchange and the publisher belongs to the same company. So, in this case, incentives are aligned.

Equation~\ref{eq::erev1} shows the publisher's expected revenue of an impression opportunity ($eRev_{pub}(\cdot)$) as a function of a vector of floors $\rho$ set by the exchange $Y$, and the set of features $X$. For instance, $X$ can include user characteristics, type of browser, hour of the day,...

{\small
\begin{eqnarray}
\label{eq::erev1}  
	eRev_{pub}(\rho|X) = eRev_{Y}(\rho|X) + Rev_{out}\!\times\! Prob(bid_i<\rho_i\; \forall i\in \mathcal{I}|X)
 \\\nonumber
\end{eqnarray}}
where the first element of the right hand side ($eRev_{Y}(\cdot)$) captures the expected revenue that the publisher would obtain if, at least, one of the submitted DSP bids in the auction conducted by \textit{Exchange Y} is above the reserve price. Each element of the vector of floors $\rho$ corresponds to the floor assigned to a specific DSP bidder ($\rho=(\rho_1,...,\rho_I)$). We denote the set of potential DSPs as $\mathcal{I}$. The second summand corresponds to the expected return from the outside option ($Rev_{out}$) if all bids submitted in \textit{Exchange Y} ($bid_i \;\forall i \in \mathcal{I}$) are below their corresponding reserve price ($\rho_i$). This happens with probability $Prob(bid_i<\rho_i \;\forall i\in \mathcal{I}|X)$. Conditional on sending the opportunity to the outside option, the expected revenue $Rev_{out}$ is constant. This assumption rule out situations where the quality of the inventory changes as a function of floors $\rho$. I.e. we could argue that the expected quality of the ad opportunities sent to the outside option is higher when floors in $\rho$ are high than when the elements in $\rho$ are low. In the Yahoo's application, we did not observe significant changes in $Rev_{out}$ as a function of floors $\rho$. So, we concluded that the assumption holds, at least for Yahoo. 

As mentioned earlier, the objective is to find the vector of floors $\rho$ that maximizes the total publisher's revenue. That is,
\begin{eqnarray}
\label{eq::erevmax}  
	\rho^{*} &=& argmax_{\rho} eRev_{pub}(\rho|X) \\\nonumber
                && s.t. \\\nonumber
                &&\rho_i=\rho_{type_j} \;if \; DSP_i\in type_j\;\forall i\in \mathcal{I},\\\nonumber
                &&\;\;\;\;\;\;\;\;\;\;\;\;\;\;\;\;\;\forall type_j \in \{regular,rebroadcaster\} \nonumber
\end{eqnarray}
where $\rho^{*}=(\rho_1^{*},\rho_2^{*},...,\rho_{I}^{*})$ is an $I$ size vector, and each element corresponds to the optimal floor assigned to a particular DSP. As mentioned before, in online advertising some DSPs can leverage their market power and impose some restrictions on how exchanges/publishers can handle the data obtained from those DSPs (e.g. submitted bids). Two typical constraints imposed by large DSPs are relevant in our case: first, exchanges/publishers cannot use intraday bid data to compute floors. Second, the floors that these DSPs face cannot differ from the floors applied to the same type of bidders. The constraint in Equation~\ref{eq::erevmax} guarantees that floors are set by type of bidder: \textit{regular} DSPs and \textit{rebroadcasters}. As a reminder, \textit{Regular} DSPs correspond to potential buyers directly bidding in the exchange on behalf of their advertising clients (e.g. TradeDesk or Yahoo DSP). On the other hand, \textit{rebroadcasters} are other exchanges bidding in the exchange that, as the name implies, "rebroadcast" advertising opportunities to their own ad exchanges and consolidate bids from multiple DSPs participating in them (e.g. OpenX or Pubmatic). \textit{Rebroadcasters} often provide additional services to help other DSPs to target users. As a result of the constraint, the bid requests sent to all \textit{regular} DSPs will have the same floor. Similarly, floors included in the bid request sent to \textit{rebroadcasters} will be equal\footnote{Under some circumstances floors may differ among bidders within the same type as a result, for example, of the existence of deals between exchanges and advertisers.}. Note that this constraint does not prevent the proposed model to be used in other domains outside the online advertising industry.

Equation~\ref{eq::erev} further develops the elements appearing in Equation~\ref{eq::erev1}, and shows how the expected returns from \textit{Exchange Y} ($eRev_{Y}(\cdot)$) and the outside option ($eRev_{out}$) depend on DSPs bidding distribution and participation rate. In order to improve readability, we dropped the set of features $X$ from the equation. 

\begin{eqnarray}
\label{eq::erev}  
	eRev_{pub}(\rho)\!\!\!\!\!\! &=&\!\!\!\!\!\! \underbrace{\sum_{i\in I}\int\left[k_i\underbrace{\int_{\rho_i}b_i\prod_{j\neq i}F_j^{k_j}(b_i|\rho_j)dF_i(b_i|\rho_i)}_{\mbox{$DSP_i$ Expected Revenue  ($eRev_{DSP_i}$)}}\right]dG_k}_{\mbox{Expected \textit{Exchange Y} Revenue ($eRev_{Y}(.)$)}} \nonumber +\\
	&& +\underbrace{Rev_{out}\int \prod_{i\in I}F_i^{k_i}(\rho_i|\rho_i)dG_k}_{\mbox{$Rev_{out} \times Prob(bid_i<\rho_i\; \forall i\in \mathcal{I})$}}
\end{eqnarray} 
where
\begin{itemize}
	\item $\rho$: vector of floors of size $I$ we want to optimize
	\item $eRev_{pub}(\rho)$: publisher's expected revenue of the inventory opportunity as a function of the vector of floors $\rho$.
 	\item $F_s(\cdot|\rho_s)\;\forall s \in \mathcal{I}$: it corresponds to the cumulative distribution function of the bids submitted by bidder $s\in \mathcal{I}$. This distribution depends on the floor set for bidder $s$ ($\rho_s$). For instance, $F_s(u|\rho_s)$ indicates the probability that the bid submitted by buyer $s$ is lower than the bid value $u$. $F_s(\cdot|\rho_s)$ is a function that must be estimated before optimizing $Rev_{pub}(\rho)$. Further details can be found in the next section.
	\item $k_s \in \{0,1\}\;\forall s \in \mathcal{I}$: it is an indicator function that shows if bidder $s$ participates ($k_s=1$) or not ($k_s=0$) in the auction. The vector of bidders' participation $k=(k_1,k_2,...,k_{I})$ follows a multinomial distribution $G_k$, which must be estimated (i.e. $k \sim G_k$ ). In our implementation, we assume that the participation of each bidder follows a Bernoulli distribution with historical observed participation rate as a parameter (i.e. $k_s \sim G_{k_s} \forall s \in \mathcal{I}$, so $G_k=\prod_{i\in \mathcal{I}} G_{k_i}$ ). Note that we make the assumption that $G_k$ does not depend on floors. This assumption is motivated by our empirical application. While the winning probability for bids below floors is zero, we noticed that large DSPs may still bid below the floor. 
 	\item $Rev_{out}$: publisher's expected revenue if the inventory opportunity is sent to another exchange (outside option).
\end{itemize}

In the first summand in the right hand side of Equation~\ref{eq::erev}, the expression denoted as $eRev_{DSP_i}$ captures the expected revenue coming from DSP $i$ conditional on participating in the auction ($k_i=1$), and given the set of competing participants ($k_j\in \{0,1\}\; \forall j\neq i$). To win the auction, DSP $i$'s bid ($b_i$) must be both greater than the submitted competing bids and higher than the assigned floor $\rho_i$. Conditional on bid $b_i$, DSP $i$ wins against the rest of bidders with probability $\prod_{j\neq i}F_j^{k_j}(b_i|\rho_j)$. Floors are computed before the auction takes place. So, ex-ante, $b_i$ is unknown to the exchange. That is why we compute the expected bid of DSP $i$ using the bid distribution $F_i(\cdot)$ over the interval $[\rho_i,\infty]$. The total expected return from the internal auction ($eRev_{Y}(\cdot)$) is the sum of returns coming from each of the potential bidders in $\mathcal{I}$.

The second summand in the right hand side of Equation~\ref{eq::erev} captures the expected revenue from the outside option when \textit{Exchange Y} does not have a winner when conducting the auction. This happens when all participant DSPs submit a bid below their corresponding floor, and with probability $\int \prod_{i\in \mathcal{I}}F_i^{k_i}(\rho_i|\rho_i)dG_k$.  

As we will note in detail in the model's implementation section, before finding the floors that maximize the expected publisher's revenue (solving Equation~\ref{eq::erevmax}), we need estimates for bids and participation distributions ($F_i()$ and $G_k$ respectively). We assume that these distributions follow a parametric form, and we rely on an experimental bucket with randomized floors to obtain estimates of the parameters that characterize the distributions.

\section{Model's Implementation at Yahoo}
This section highlights the model implementation described above on display and video advertising inventory appearing in Yahoo, which includes AOL sites.

\begin{figure}[htbp]
        \captionsetup{justification=centering}
\caption{Bid Distributions for Low/High\newline Reserve Prices and QTE}
\hfill
%\subfigure[Low Reserve Price]{\includegraphics[width=1.5in]{figures/Placement767521Date20230223CriteoFloor015BucketOptimalFloorDisabled.png}}
\subfigure[Low Reserve Price]{\includegraphics[width=1.5in]{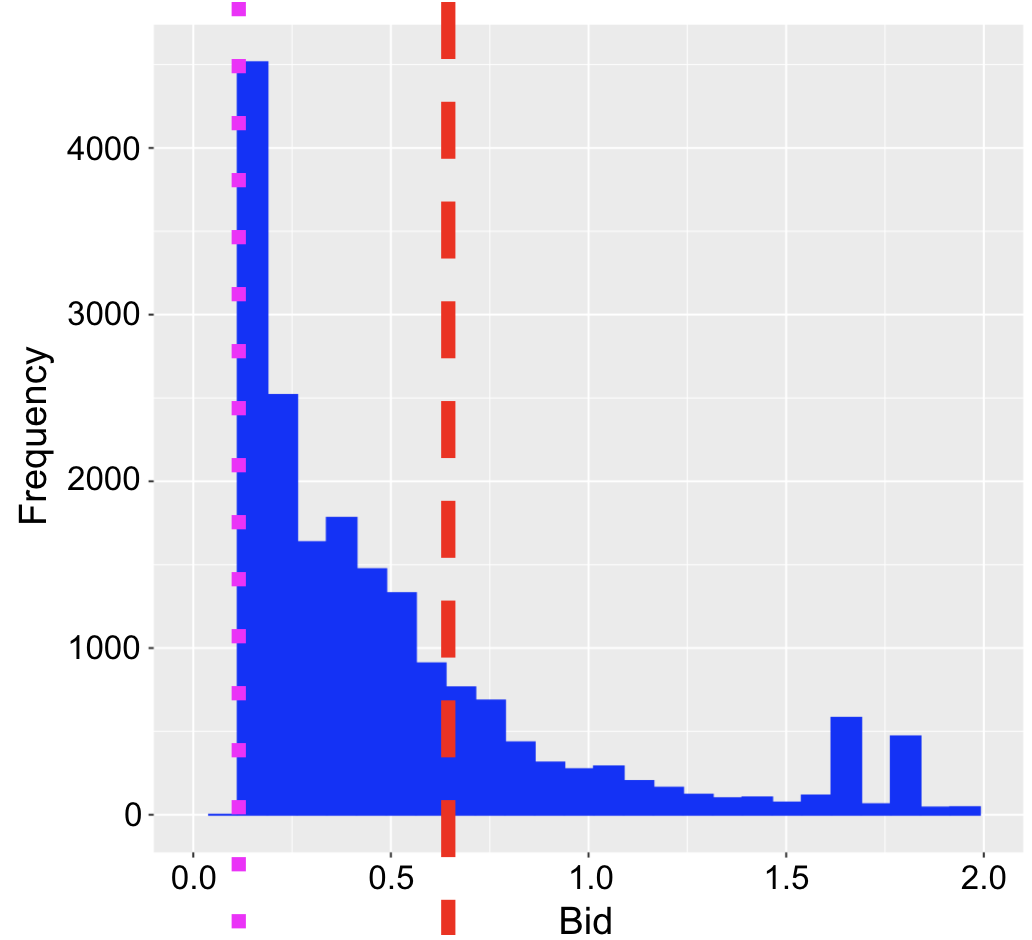}}
\hfill
%\subfigure[High Reserve Price]{\includegraphics[width=1.5in]{figures/Placement767521Date20230223CriteoFloor067BucketOptimalFloorProdModel.png} }
\subfigure[High Reserve Price]{\includegraphics[width=1.5in]{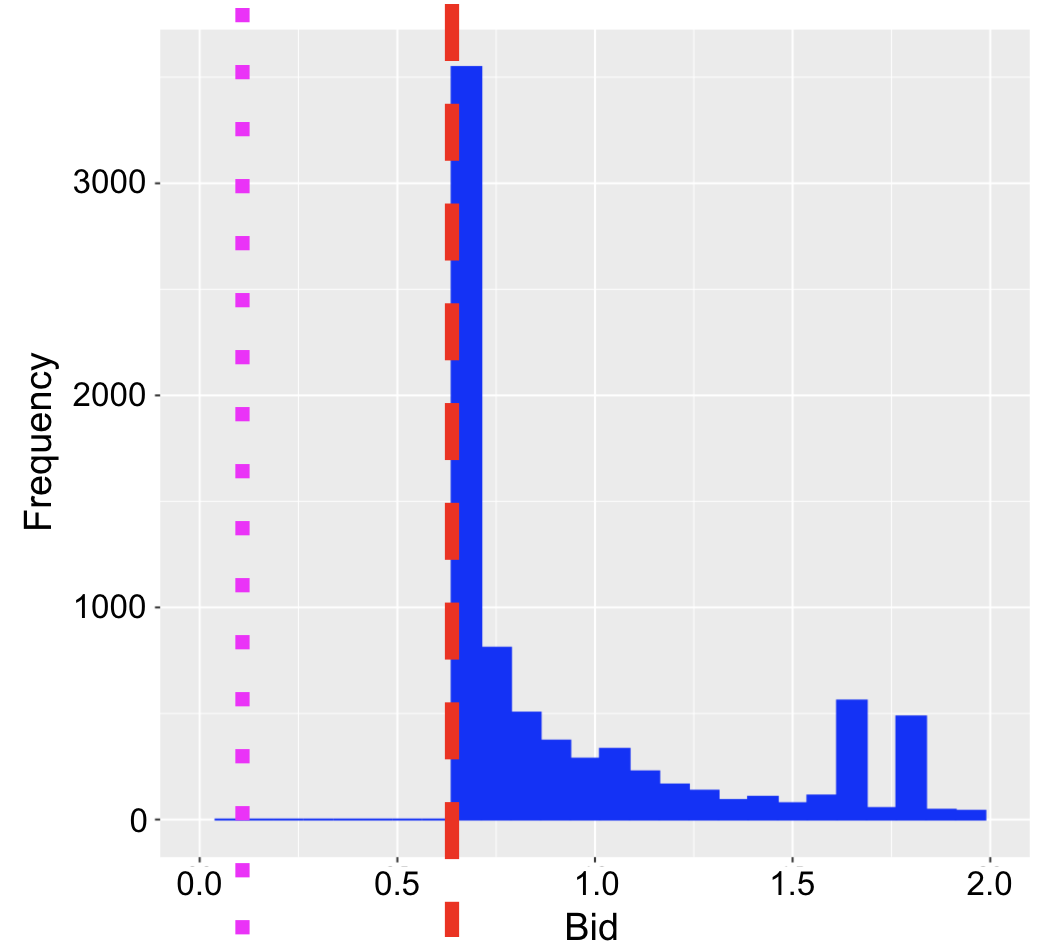} }
\hfill
%\subfigure[QTE]{\includegraphics[width=1.5in]{figures/QTEPlacement767521Date20230223CriteoFloor067.png}}
%\subfigure[QTE]{\includegraphics[width=1.5in]{figures/QTEcriteoclean.png}}
\subfigure[QTE]{\includegraphics[width=1.5in]{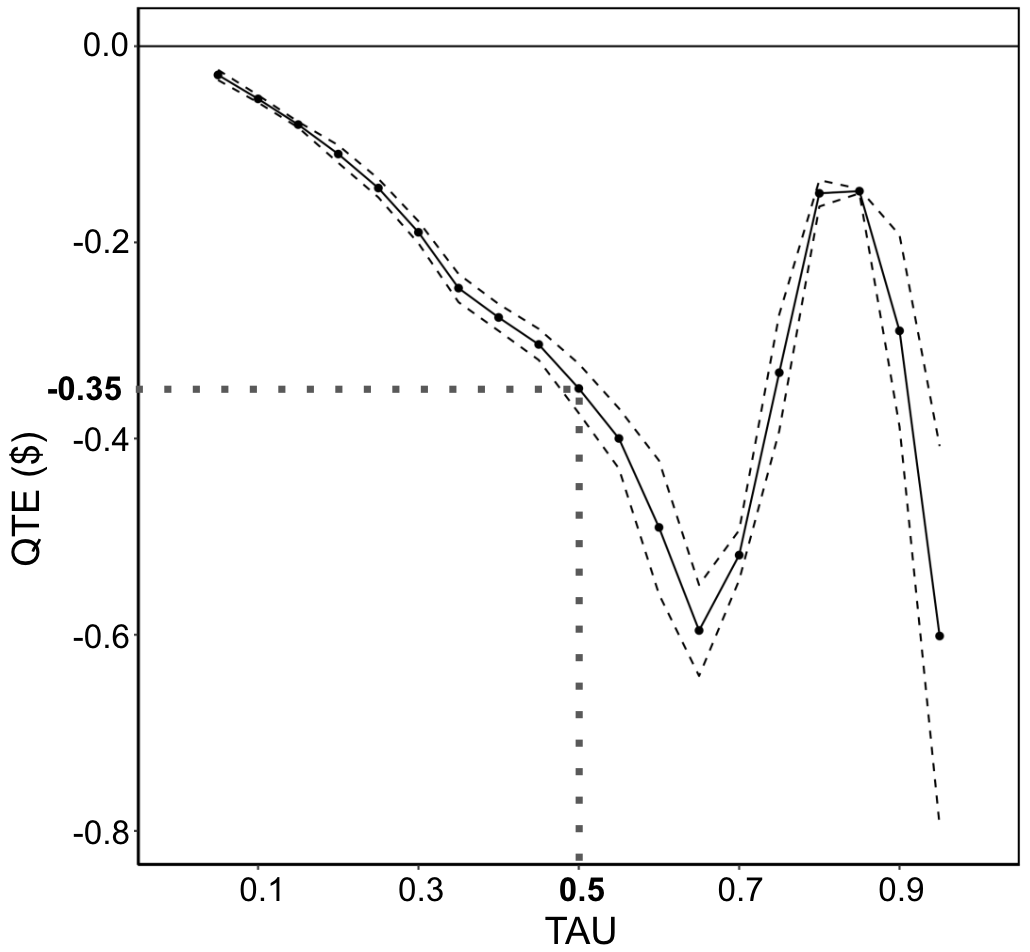}}

\hfill\\
\label{biddist_3}
    \small
    \bfseries Note: \normalfont Figure shows the bid distributions of a particular DSP bidding on a particular Yahoo's ad inventory when: (a) the floor is low, and (b) the floor is high. (c) captures differences between both distributions for bids above the high floor using QTEs.
    \medskip
\end{figure}

Consistent with the expected bidder behavior described in the introduction, most of the DSPs react to the floor enclosed in the bid request. As an illustration, Figure~\ref{biddist_3} displays the bidding distributions of a DSP bidding on a particular Yahoo's ad location, and facing different floor values. The data is generated using experimental buckets with the same percentage of traffic in bid requests. DSP, ad location, and Yahoo property have been anonymized for confidentiality reasons. Figure~\ref{biddist_3}(a) shows the distribution of bid responses of the DSP when the floor observed by the DSP is low (pink dotted line). Figure~\ref{biddist_3}(b) corresponds to the distribution when the floor enclosed in the bid request is high (red dashed line), and Figure~\ref{biddist_3}(c) captures the difference between both distributions for bid values above the high reserve price using Quantile Treatment Effects (QTE). The QTE is a statistical tool to analyze differences between distributions using quantiles (see \cite{firpo07} and \cite{callaway23}). The quantile treatment effects (QTEs) are defined as the difference between the quantiles of the treated and untreated outcome distributions. In our application, the treated distribution corresponds to the distribution of bids above the high floor when the floor enclosed in the bid request is high (Figure~\ref{biddist_3}(b)). The untreated distribution corresponds to the distribution of bids above the high floor when the floor enclosed in the bid request is low (Figure~\ref{biddist_3}(a)). In Figure~\ref{biddist_3}(c), each point within the continuous line  corresponds to the difference between the bids coming from high and low floor distributions at a particular quantile (TAU). The dashed lines correspond to the 95\% confidence interval thresholds. For instance, if we look at the difference between the median bids above the high floor in both distributions (TAU=0.5), Figure~\ref{biddist_3}(c) suggests that the median bid in the high floor distribution is 35 cents lower than the median bid in the low floor distribution, and this difference is statistically significant. If the DSP did not react to the floors enclosed in the bid request, both distributions would be equal and the QTE line would lie around zero. However, as the reader can easily identify, the QTE is significantly different from zero for any quantile level. This result indicates that both distributions are different, in line with the behavior depicted in the introduction (Figure~\ref{fig::biddingdistribution}).

 In the implementation, we compute floors at placement level. At Yahoo, placements are the most granular piece of supply inventory, and indicates the ad location in the page, the site, the ad size, and the format. For instance, a placement may correspond to a 300x250 pixels display ad, located in the upper-right corner of Yahoo Mail. For each placement we will have two floors: one for $regular$ DSPs and the other targeting $rebroadcasters$.

\subsection{Finding the optimal floors}
In this section we include a few remarks about how to compute the vector of optimal floors $\rho^*$. To solve the optimization problem in Equation~\ref{eq::erevmax}, we first need to have an estimate of the bidding distribution of DSPs as a function of floors ($F_i(\rho|X)$ $\forall i \in \mathcal{I}$), and corresponding participation rate ($G_{k_i}(\rho|X)$ $\forall i \in \mathcal{I}$).

\begin{figure}[htbp]
	\caption{Training and Output Components}
	\label{fig::components}
	\centering
	\includegraphics[width=.48\textwidth]{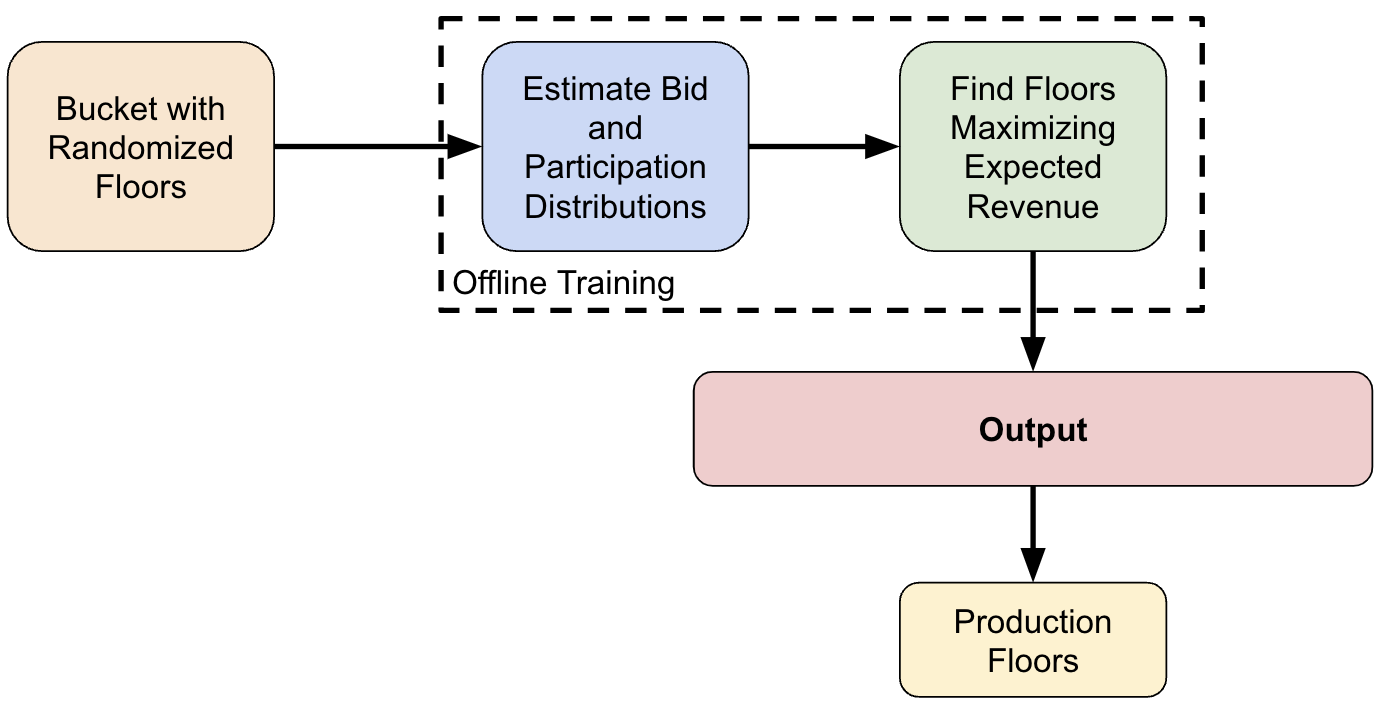}

 	%\medskip
    \small
    \bfseries Note: \normalfont Figure illustrates the different training and output components
    %\medskip
\end{figure}

Figure~\ref{fig::components} displays the different training and output components included in the pipeline:
\begin{enumerate}
    \item \bfseries Bucket with randomized floors: \normalfont We need exogenous variation in floors in order to obtain unbiased estimates of the effect of floors on the bidding distribution of each DSP. We use a 1\% bucket with randomized reserve prices, which allows us to observe how the bidding behavior of DSPs changes as a function of the floor passed in the bid request.  We use the 95 percentile of historical bid values as an upper bound to avoid having unnecessary high floors in our randomized bucket. We use 7 days of randomized floors data to train our models.
    \item \bfseries Estimation of bidding and participation distributions $F_i(\rho|X)$ and $G_k(X)$: \normalfont Before being able to find the vector of optimal floors ($\rho*$), we need to estimate the cumulative distribution function of the bids submitted by DSPs as a function of floors ($F_i(\rho|X)$). Similarly, we need to estimate the auction participation of DSPs ($G_k(X)$). 
    
    We assume that the bid distribution of each buyer follows a Weibull Distribution, which is characterized by two parameters: shape ($\mu_{shape}>0$) and scale ($\mu_{scale}>0$). The shape parameter is a linear function of floors and is parameterized as follows,

    \begin{eqnarray}
    \label{eq::weibull}
        &&\mu_{i,shape}=EXP(\theta_{i,0} + \theta_{i,1}*\rho_i + \theta_{i,2}*\rho_i^2) \\\nonumber
        &&\mu_{i,scale}
    \end{eqnarray} 
        
    $\theta_i = \{\mu_{i,scale},\theta_{i,0},\theta_{i,1},\theta_{i,2}\}$ corresponds to the set of parameters defining the bidding distribution of a particular DSP $i$ as a function of floors, and must be estimated. The observables (bid and floors) are obtained from the bucket with randomized floors, and the parameters are estimated independently for each DSP using maximum log likelihood. 
    
    \begin{eqnarray}
    \label{eq::likelihood}
        &&\hat{\theta}_i= argmax_{\theta_i} \frac{1}{T_i}\sum_tlog f_{i,t}(bid|\theta_i,\rho_{i,t})\;\; \forall i\in\mathcal{I}
    \end{eqnarray}   
    where $f_i()$ corresponds to the Weibull probability distribution function of bids submitted by DSP $i$, and $T_i$ corresponds to the total number of bidder $i$ observations in the bucket.

    As discussed in the model section, we assume that the participation of each bidder follows a Bernoulli distribution with historical observed participation rate as a parameter (i.e. $k_s \sim G_{k_s} \forall s \in \mathcal{I}$ ).

    \item \bfseries Find the optimal vector of floors $\rho^{*}$: \normalfont Given the estimates of the bidding and participation distributions, we solve the maximization problem stated in Equation~\ref{eq::erevmax}, and find the vector of reserve prices ($\rho$) that maximizes the expected revenue of the inventory opportunity ($eRev_{pub}(\rho)$). As noted previously, we had to introduce constraints to guarantee that floors are set by type of DSP: $regular$ and $rebroadcasters$.
    
    Floors are computed daily. Figure~
    \ref{fig::floortrend} displays the optimal floors of a sample of placements (AAAA, BBBB, CCCC) by type of bidder ($regular$ DSPs and $rebroadcasters$) from November 8th 2022 until November 22nd 2022. Placement identifiers have been anonymized for confidentiality reasons. The Figure shows that the obtained floors are pretty stable. This suggests that the optimization solver is able to find the global maximum when solving Equation~\ref{eq::erevmax}.

\begin{figure}[htbp]
	\caption{  Stability of Optimal Floor Outputs}
	\label{fig::floortrend}
	\centering
        \includegraphics[width=.46\textwidth]
        {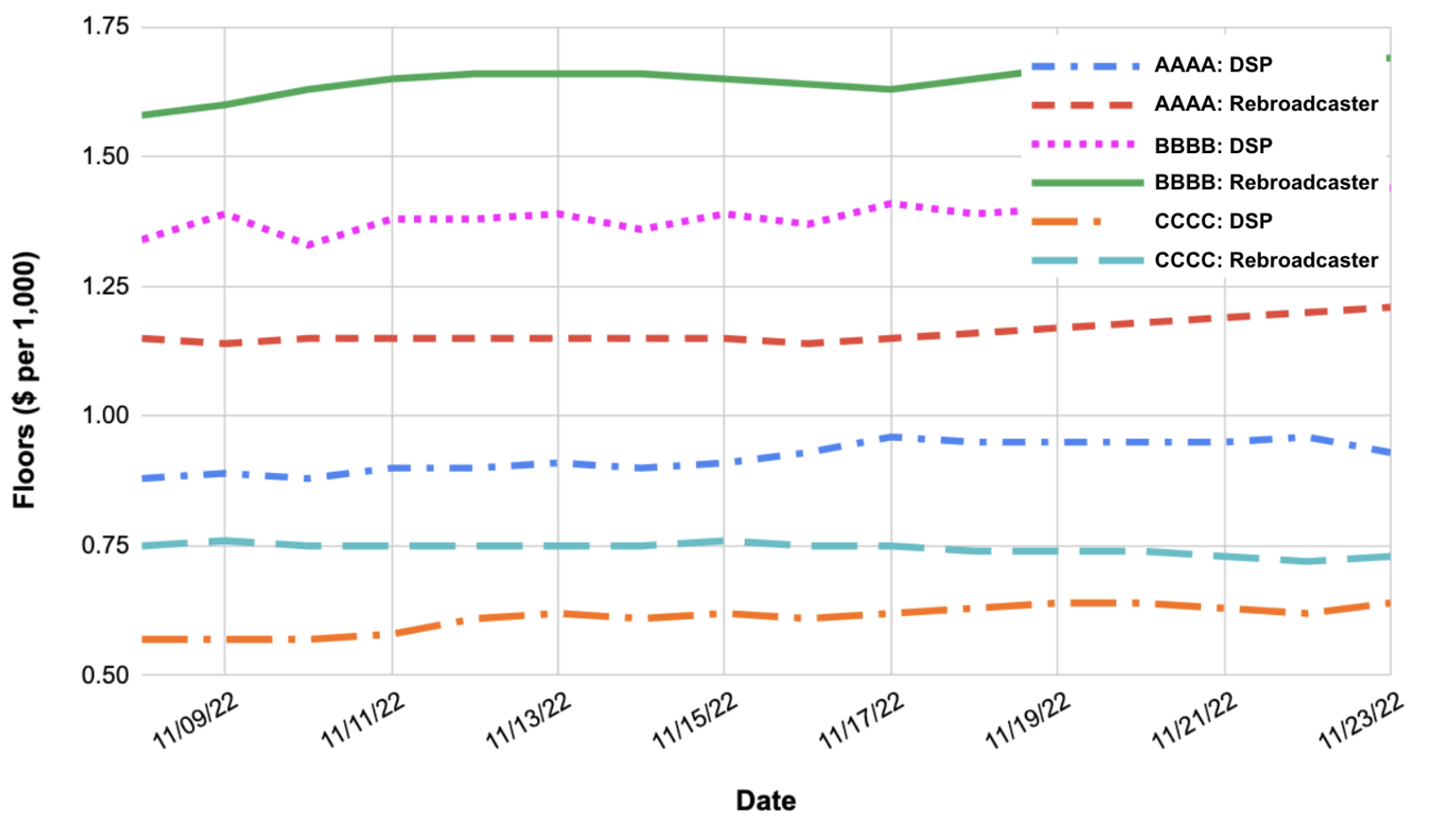}%{figures/floorstabilityanonymized.png}
    
    % \medskip
    \small
    \bfseries Note: \normalfont Figure shows the stability of optimal floor outputs for a sample of placements by type of bidder from November 8th 2022 until November 22nd 2022.
    \medskip
\end{figure}
    
    \item \bfseries Output: \normalfont The output corresponds to a csv file where each row contains the features we are considering and the floors associated to them. Our current implementation supports placement, site and publisher ids. As a result, each row of the file contains the corresponding placement, site and publisher ids and the associated floors for $regular$ DSPs and $rebroadcasters$. As an illustration, Table~\ref{table::samplefloors} shows a sample of the output generated last November 8th, 2022. It contains three placement ids and corresponding floors for $regular$ DSPs and $rebroadcasters$.  In the table, we can see that for placement ID AAAA we attach a \$0.88 floor value in bid requests sent to $regular$ DSPs and \$1.15 to $rebroadcasters$.
    
    The output file will be validated making sure that results are consistent with expectations. We use Tukey's Method to detect potential outliers in the results. If the output file passess the validation, it will be deployed in production.
    \item \bfseries Floors in production: \normalfont Every time there is an ad opportunity, the system looks at the deployed output and attaches the corresponding floor to the bid request.
\end{enumerate}

{%\small
\begin{table}[htbp]
	\centering
	\setlength{\abovecaptionskip}{2pt}
	\caption{Sample of output floors - November 8th 2022}
	\label{table::samplefloors}
	%\normalsize
        \small
	\begin{threeparttable}
		\begin{tabular}{|c|c|c|c|c|} %\hline
            \bfseries Publisher & \bfseries Site & \bfseries Placement & \bfseries Regular & \bfseries Rebroadcaster	\\ \hline
             ABCD & KKKK & AAAA & \$0.88 & \$1.15\\ \hline
             ABCD & KKKK & BBBB & \$1.34 & \$1.58\\ \hline
             ABCD & HHHH & CCCC & \$0.57 & \$0.75  \\ \hline         
            \bfseries \vdots &	\vdots &	\vdots & \vdots & \vdots	\\ \hline	
		\end{tabular}
	\end{threeparttable}

	\medskip
    \small
    \bfseries Note: \normalfont Table contains a sample of the floor output\\ generated on November 8th
\end{table}
}\normalsize
We tested the model using a log-normal distribution instead of the Weibull, and obtained similar results. However, the Weibull distribution led to a faster convergence and less errors when solving the optimization problem to find the floors.

\subsection{Results}

The outlined methodology has been implemented at Yahoo with remarkable results.
In June 2021, we deployed the model in production on display ad inventory in Yahoo websites located in North-America. Afterwards, we rolled out the feature in other markets, and in October 2022 we started optimizing floors on Yahoo video ad inventory. This product significantly improves the previous approach to set floors: a manual process where floors were equal to the expected returns from the outside option, without taking into account the impact of floors on the bidding behavior of DSPs. The annualized incremental revenue is estimated at +1.3\% on Yahoo display inventory and +2.5\% on video ad inventory. These are non-negligible numbers in a multi-million Yahoo ad business.

As discussed in the model section, Yahoo owns the main ad exchange that allocates and prices ads (denoted as $YahooX$). On the other hand, Yahoo uses Google Ad Exchange ($ADX$) as the outside option when $YahooX$ is not able to find an ad for the impression opportunity.

A lot of enhancements have been introduced since the first launch in June 2021. To understand how the Floor feature impacts the marketplace, we are going to focus on Yahoo display inventory and results from the 3rd quarter of 2022. A period without big deployments. 

\begin{figure}[htbp]
	\caption{Revenue Lift (\%) Yahoo Display Inventory - Q3 2022}
	\label{fig::timeseries}
	\centering
 	\includegraphics[width=.45\textwidth]{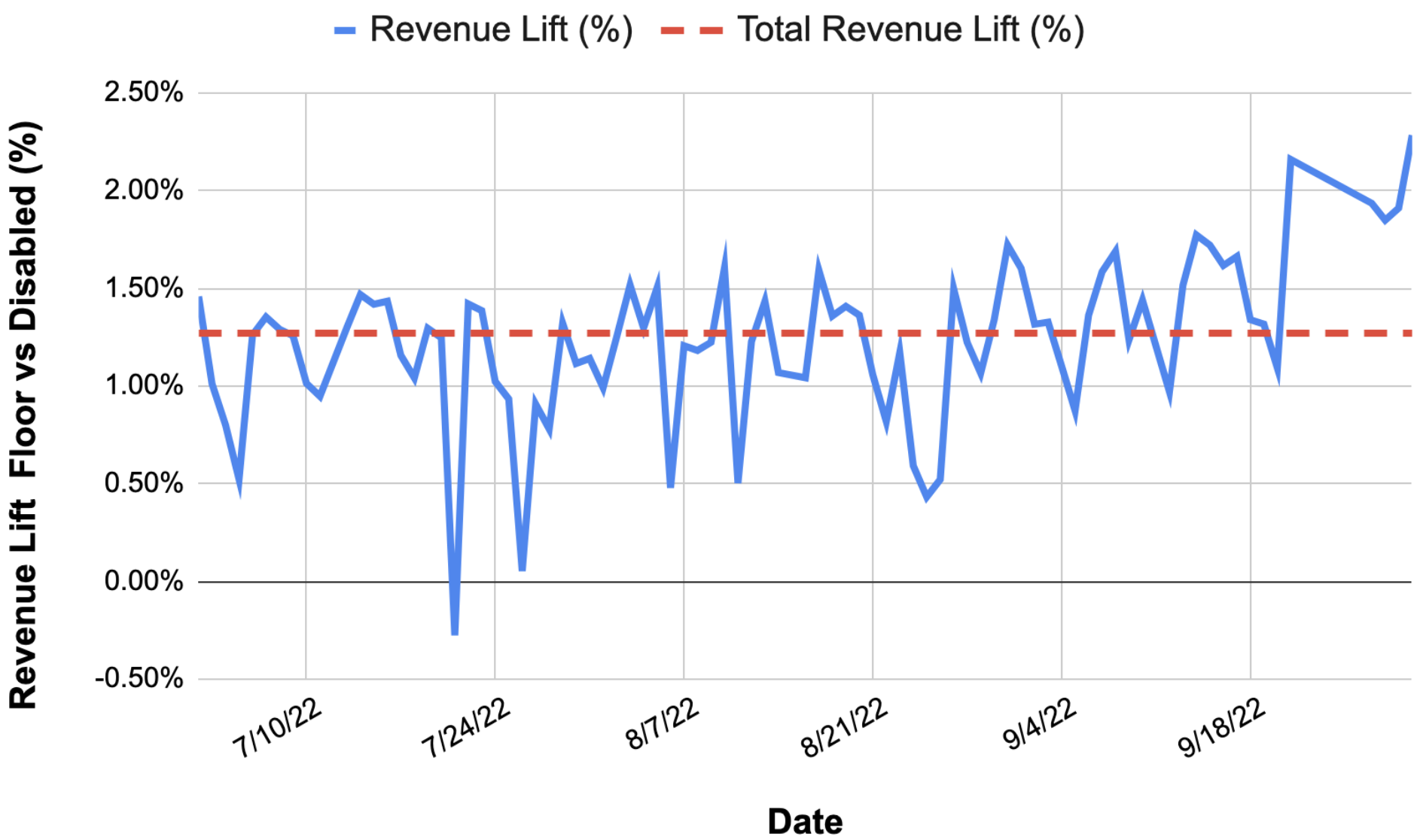}
    \medskip
    \small

    \bfseries Note: \normalfont Figure captures the revenue lift impact, in \%, of the floor feature on Yahoo display inventory during Q3 2022.
    \medskip
\end{figure}

Figure~\ref{fig::timeseries} displays the estimated revenue impact, in \%, of the Floor feature during the 3rd quarter of 2022. The x-axis displays the date and the y-axis captures the difference between the estimated revenue generated by the algorithmic Floors and the revenue we would have had if we continued setting floors manually. We can see that the daily revenue lift is always positive and, on average, equal to 
 1.27\%. To assess the impact, we compare the performance of a bucket with 5\% traffic that mimics production ( \textit{Floor Bucket} \normalfont ) with a 5\% bucket with disabled Floors ( \textit{Disabled Bucket} \normalfont), which corresponds to the previous manual approach to set reserve prices.

\begin{table}[htbp]
	\centering
	\setlength{\abovecaptionskip}{2pt}
        \captionsetup{justification=centering}
	\caption{Impact of the Floor Feature on Yahoo\newline Display Inventory - Q3 2022}
	\label{table::impactoonar}
	\normalsize
	\begin{threeparttable}
		\begin{tabular}{|c|c|} %\hline
         & \bfseries Lift (\%) \\ \hline
        \bfseries Total Revenue: YahooX + ADX &	1.27\%	\\ \hline
         YahooX Revenue 	 &0.41\% \\ \hline
         ADX Revenue  &	64.48\% 	\\ \hline \hline
        \bfseries Total Impressions: YahooX + ADX &		-1.33\%	\\ \hline
         YahooX Impressions &		-13.86\% \\ \hline
         ADX Impressions  &		39.28\% \\ \hline

         ADX Impressions Share \%	& 	41.17\%  \\ \hline
        \hline
        \bfseries Total eCPM: YahooX + ADX &	2.64\%\\ \hline
         YahooX eCPM &		16.57\%	\\ \hline
        \hline
	\end{tabular}
	\end{threeparttable} 
    
    \medskip
    \small
    \bfseries Note: \normalfont Table summarizes the impact of the floor feature on Yahoo display inventory during Q3 2022.
    \medskip
\end{table}

Table~\ref{table::impactoonar} summarizes results for Yahoo display inventory during the 3rd quarter of 2022. In Q3 2022, the floor initiative generated +1.27\% in incremental revenue. If we look at the revenue impact by origin, the return coming from clearing auctions in Yahoo's internal exchange ($YahooX$) increased by 0.41\%, while the revenue from the Google $ADX$ outside option solution increased by 64.48\%. The dynamics explaining the revenue impact on both exchanges, $YahooX$ and $ADX$, are different. If the reader recalls Figure~\ref{fig::biddingdistribution}, raising floors induces DSPs to increase their bids, shifting the bidding distribution to the right. However, bids will never be higher than the valuation for the inventory opportunity. As a result, the number of bids above the floor declines but at the same time the clearing price of the impressions rises. This is what happens in $YahooX$: there is a 13.86\% drop in the number of impressions served by Yahoo's exchange and a 16.57\% increase in eCPM, defined as revenue per 1,000 requests. The increase in revenue coming from the rise in the number of high bids compensates the lost associated with the drop in the number of impressions, leading to an increase in $YahooX$ revenue (+0.41\%). The drop in $YahooX$ impressions explains the increase in the number of impressions served by $ADX$ (+39.28\%) and, subsequently, the increase in revenue (+64.48\%). Another important metric is the variable \textit{ADX Impressions Share \%}, which measures the ratio between the number of $ADX$ impressions and the total number of impressions. This ratio helps to understand how much Yahoo relies on Google $ADX$ in terms of monetization. The \textit{ADX Impressions Share \%} increased considerably (+41.17\%), triggering some discussions among stakeholders about the need (or not) of constraining the number of requests sent to $ADX$.

\begin{table}[htbp]
	\centering
	\setlength{\abovecaptionskip}{2pt}
        \captionsetup{justification=centering}
	\caption{Total Impact of the Floor Feature on Yahoo\newline Display Inventory by Site - Q3 2022}
	\label{table::impactsiteall}
	%\normalsize
        \small
	%\begin{threeparttable}
		\begin{tabular}{|c|c|c|c|} %\hline
         & \multicolumn{3}{c}{\bfseries Total: YahooX + Google ADX}\\
        \bfseries Site Name  & \bfseries Revenue Lift &	\bfseries eCPM Lift & \bfseries Imp. Lift\\ \hline
        \bfseries Yahoo NAR	Mail &	1.36\% &	2.69\% &	-1.41\% \\ \hline
        \bfseries Yahoo NAR	Home &	1.07\% &	1.65\%&	-0.70\%  \\\hline
        \bfseries AOL NAR	Webmail &	1.22\% &	1.08\%&	0.06\% \\ \hline
        \bfseries Yahoo NAR	News &	1.35\% &	4.20\%&	-2.94\% \\ \hline
        \bfseries Yahoo NAR	Finance &	1.31\% &	3.57\%&	-2.33\%\\ \hline
 	\end{tabular}
	%\end{threeparttable} 
    
    \medskip
    \small
    \bfseries Note: \normalfont Table summarizes the total impact of the floor feature on Yahoo display inventory by site during Q3 2022.
    \medskip
\end{table}\normalsize

\normalsize
Table~\ref{table::impactsiteall} shows the impact of the Floor feature on the most popular Yahoo sites in terms of revenue (all of them located in North America (NAR)). Columns 2-4 capture changes in revenue, ecpm and impressions on the overall Yahoo marketplace ($YahooX$ plus $ADX$). The total revenue lift on all displayed properties is greater than 1.0\%. As expected, eCPMs increased for all sites, specially in News and Finance Yahoo sites (+4.20\%, and +3.57\% respectively). 

\begin{table}[htbp]
	\centering
	\setlength{\abovecaptionskip}{2pt}
        \captionsetup{justification=centering}
	\caption{Impact of the Floor Feature on $YahooX$\newline Display Inventory by Site - Q3 2022}
	\label{table::impactsiteX}
	%\normalsize
        \small
	\begin{threeparttable}
		\begin{tabular}{|c|c|c|c|} %\hline
         & \multicolumn{3}{c}{\bfseries YahooX}\\
        \bfseries Site Name  & \bfseries Revenue Lift &	\bfseries eCPM Lift & \bfseries Imp. Lift\\ \hline
        \bfseries Yahoo NAR	Mail &	1.09\% &	8.26\% &	-6.64\% \\ \hline
        \bfseries Yahoo NAR	Home &	0.81\% &	5.53\% &	-4.17\% \\\hline
        \bfseries AOL NAR	Webmail &	1.21\% &	3.81\% &	-2.34\%\\ \hline
        \bfseries Yahoo NAR	News &	-0.16\% &	27.20\% &	-21.15\% \\ \hline
        \bfseries Yahoo NAR	Finance &	-0.26\% &	23.46\% &	-18.93\% \\ \hline
 	\end{tabular}
	\end{threeparttable} 
    
    \medskip
    \small
    \bfseries Note: \normalfont Table summarizes the impact of the floor feature on $YahooX$ display inventory by site during Q3 2022.
    \medskip
\end{table}\normalsize

Table~\ref{table::impactsiteX} shows the impact  of the Floor feature on Yahoo's internal exchange ($YahooX$) by the most popular Yahoo sites in terms of revenue. We can see that the revenue lift is still positive for all sites except for Yahoo News and Finance (-0.16\% and -0.26\% respectively). The impact on eCPM and Impressions is pretty significant. For instance, the eCPM of impressions served by $YahooX$ on Yahoo Mail increased by 8.26\%, while the number of impressions decreased by 6.64\%.

\section{Conclusions}
In this paper we present a model to set floors in first price auctions, and discuss the impact of its implementation on Yahoo sites. The model captures important characteristics of the online advertising industry, such as the existence of an outside option and contractual constraints imposed by some bidders. Our solution induces bidders to change their bidding behavior as a response to the floors enclosed in the bid request, helping online publishers to increase their ad revenue. The outlined model has been implemented at Yahoo with remarkable results. The annualized incremental revenue is estimated at +1.3\% on Yahoo display inventory, and +2.5\% on video ad inventory. 

The paper does not address potential long term effects of the proposed model. Demand side platforms bid on behalf of their advertiser clients. Advertisers rely on DSPs to deliver their budgets under some goal constraints. As pointed out in the results section, the floor algorithm substantially increases the cost per impression of ads served by Yahoo's exchange. This cost surge may impact the long term behavior of DSPs on Yahoo. The employed AB test framework to assess the impact of the floor feature is not able to measure the potential future consequences.

\clearpage
\thispagestyle{empty}
\bibliographystyle{ACM-Reference-Format}
\bibliography{miguel} 

\end{document}